%Paper: hep-th/9506063
%From: li@het.brown.edu (Miao Li)
%Date: Fri, 9 Jun 95 14:14:50 EDT
%Date (revised): Fri, 9 Jun 95 14:47:27 EDT

% the following is to incorporate them in the body of the text,
% using epsf.
\let\includefigures=\iftrue
%
% activate this if you don't have epsf.
\let\includefigures=\iffalse
%
% the following is to use blackboard bold fonts --
\let\useblackboard=\iftrue
%
% activate this if you don't have them.
%\let\useblackboard=\iffalse
%
%
\input harvmac.tex
%\includefigures
\message{If you do not have epsf.tex (to include figures),}
\message{change the option at the top of the tex file.}
\input epsf
\epsfverbosetrue
%\epsfclipon
\def\fig#1#2{\topinsert\epsffile{#1}\noindent{#2}\endinsert}
%\else
\def\fig#1#2{}
%\fi
%
\def\Title#1#2{\rightline{#1}
\ifx\answ\bigans\nopagenumbers\pageno0\vskip1in%
\baselineskip 15pt plus 1pt minus 1pt
\else%\special{papersize=11in,8.5in}%
\def\listrefs{\footatend\vskip 1in\immediate\closeout\rfile\writestoppt
\baselineskip=14pt\centerline{{\bf References}}\bigskip{\frenchspacing%
\parindent=20pt\escapechar=` \input
refs.tmp\vfill\eject}\nonfrenchspacing}
\pageno1\vskip.8in\fi \centerline{\titlefont #2}\vskip .5in}

\ifx\answ\bigans\def\tcbreak#1{}\else\def\tcbreak#1{\cr&{#1}}\fi
\useblackboard
\message{If you do not have msbm (blackboard bold) fonts,}
\message{change the option at the top of the tex file.}
\font\blackboard=msbm10 scaled \magstep1
\font\blackboards=msbm7
\font\blackboardss=msbm5
\newfam\black
\textfont\black=\blackboard
\scriptfont\black=\blackboards
\scriptscriptfont\black=\blackboardss

\else

\fi
% *************************************
%\draft
%
\def\yboxit#1#2{\vbox{\hrule height #1 \hbox{\vrule width #1
\vbox{#2}\vrule width #1 }\hrule height #1 }}
\def\fillbox#1{\hbox to #1{\vbox to #1{\vfil}\hfil}}
\def\ybox{{\lower 1.3pt \yboxit{0.4pt}{\fillbox{8pt}}\hskip-0.2pt}}
\def\comments#1{}

\def\p{\partial}

\def\tr{{\rm tr\ }}

\Title{\vbox{\baselineskip12pt
\hfill{\vbox{
\hbox{BROWN-HET-996\hfil}
\hbox{hep-th/9506063}}}}}
{\vbox{\centerline{Quenched Two Dimensional Supersymmetric }
\vskip20pt
\centerline{Yang-Mills Theory}}}
\centerline{Miao Li}
\smallskip
\centerline{Department of Physics}
\centerline{Brown University}
\centerline{Providence, RI 02912}
\centerline{\tt li@het.brown.edu}
\bigskip
\noindent

By studying the pure Yang-Mills theory on a circle,
as well as an adjoint scalar coupled to the gauge field on a circle, we
propose a quenching prescription in which the combination of the spatial
component of the gauge field and $P$ is treated as a dynamic variable.
Averaging over momentum is not necessary, therefore the usual ultraviolet
cut-off is eliminated. We then apply this prescription to study the
large $N$ two dimensional supersymmetric gauge theory. An one dimensional
supersymmetric matrix model is obtained. It is not known whether this
model can be solved exactly. However, an extended model with one more
complex fermion is exactly solvable, with $N=1$ supersymmetry as
Parisi-Sourlas supersymmetry. The exact solvability may have some
implications for the $N=1$ quenched model.

\Date{June 1995}
\nref\miao{M.~Li, ``Large $N$ Solution of the 2D Supersymmetric Yang-Mills
Theory'', BROWN-HET-989, hep-th/9503033, to appear in Nucl. Phys. B.}
\nref\tit{Y.~Matsumura, N.~Sakai and T.~Sakai, ``Mass spectrum of
supersymmetric
Yang-Mills theories in (1+1)-dimensions'', TIT-HEP-290, hep-th/9504150.}
\nref\sw{N.~Seiberg and E.~Witten, Nucl. Phys. B426 (1994) 19,
hep-th/9407087.}
\nref\iss{K.~Intriligator, N.~Seiberg and S.~Shenker, Phys. Lett. B342
(1995) 152, hep-th/9410203.}
\nref\seiberg{N.~Seiberg, Nucl. Phys. B435 (1995) 129, hep-th/9411149.}
\nref\af{P.~C.~Argyres and A.~E.~Faraggi, Phys. Rev. Lett. 73 (1994) 19,
hep-th/9411057; A.~Klemm, W.~Lerche, S.~Theisen and S.~Yankielowicz,
Phys. Lett. B344 (1995) 169, hep-th/9411048.}
\nref\ds{M.~R.~Douglas and S.~H.~Shenker, `` Dynamics of SU(N) Supersymmetric
Gauge Theory'', hep-th/9503163.}
\nref\ek{T.~Eguchi and H.~Kawai, Phys. Rev. Lett. 48 (1982) 1063.}
\nref\bhn{G.~Bhanot,U.~M.~Heller and H.~Neuberger, Phys. Lett. B113
(1982) 47.}
\nref\parisi{G.~Parisi, Phys. Lett. B112 (1982) 463.}
\nref\gk{D.~J.~Gross and Y.~Kitazawa, Nucl. Phys. B206 (1982) 440;
S.~R.~Das and S.~Wadia, Phys. Lett. B117 (1982) 228.}
\nref\nkw{H.~Neuberger, Phys. Lett. B119 (1982) 179; Y.~Kitazawa and
S.~R.~Wadia, Phys. Lett. B120 (1983) 377.}
\nref\ps{G. Parisi and N.~Sourlas, Phys. Rev. Lett. 43 (1979) 744; Nucl. Phys.
B206 (1982) 321.}
\nref\antal{A.~Jevicki and H.~Levine, Ann. Phys. 136 (1981) 113.}
\nref\sumit{S.~Das, Rev. Mod. Phys. 59 (1987) 235.}
\nref\js{A. Jevicki and B.~Sakita, Phys. Rev. D22 (1980) 467.}
\nref\dmp{M.~R.~Douglas, hep-th/9303159; J.~A.~Minahan and
A.~P.~Polychronakos, Phys. Lett.B312 (1993)155, hep-th/9303153.}
\nref\wb{J.~Wess and J.~Bagger, ``Supersymmetry and Supergravity,'' Princeton
University Press, Princeton (1983).}
\nref\shif{F. Lenz, M~Shifman and M.~Thies, TPI-MINN-94/34-T, hep-th/9412113.}
\nref\nicolai{H. Nicolai, Phys. Lett. B89 (1980) 341; Nucl. Phys. B176
(1980) 419.}

\newsec{Introduction}
The two dimensional supersymmetric Yang-Mills theory may turn out to be a
two dimensional matrix model which can be solved
in the large $N$ limit. A set of loop equations are solved in \miao\
with a certain assumption, its spectrum is studied numerically in \tit\
in the light-cone formalism. There seems to be at least two major motivations
for studying supersymmetric gauge theories in various dimensions.
The first is the hope that deeper understanding of these
theories may shed light on several longstanding open problems in
particle theory and quantum field theories, such as confinement \sw,
dynamic supersymmetry breaking \iss\ and various kinds of duality \refs{\sw,
\seiberg}. The second motivation, which is equally important if not more
important, is the search for higher dimensional solvable matrix models.
As we have learned from the study of zero and one dimensional matrix models,
string theory can be reformulated in terms of matrix models. In such models,
some spacetime dimensions are dynamically generated, and powerful
mathematical tools can be developed to calculate various physical quantities.
It may be justified to hope that string theory will be ultimately formulated
in this fashion. This latter motivation serves as the primary one for
our previous study \miao\ and the work presented in this paper.
Recently, based on the explicit solution of the low energy effective action
in the supersymmetric $SU(N)$ Yang-Mills theory in four dimensions \af,
Douglas and Shenker
extract some interesting large $N$ information about the theory \ds, and
find that in the large $N$ limit, the validity region of the low energy
solution becomes very narrow.
Because of the importance of the large $N$ solution of super-gauge theories,
it is then desirable to explore all possible valuable methods.

As long as the large $N$ problem is concerned, Eguchi and Kawai showed that
the Euclidean lattice gauge theory can be reduced to a model at a single site
with $D$ matrices \ek, here $D$ is the spacetime dimension. Unfortunately, it
was shown subsequently that the EK model suffers breakdown of global U(1)
symmetries at weak coupling in dimensions higher than two \bhn, so the EK
model is not capable of reproducing the large $N$ result of the Wilson
theory in the physical weak coupling regime. A quenched model was then
introduced in \bhn. This idea was substantially augmented by Parisi
\parisi\ and by Gross and Kitazawa \gk, these authors were able to show
that the quenched matrix model indeed correctly produces the large $N$
results for correlation functions in the weak coupling regime. We shall
follow refs.\nkw\ to work with the Hamiltonian formalism.

The quenching prescription proposed in this paper is slightly different
from the conventional one. Instead of first calculating quantities with
fixed quenching momentum matrix, and then averaging over momenta, we propose
to consider the combination of the momentum matrix with the spatial
component of the gauge field as a dynamic variable. The motivation for
doing this is from consideration of the pure Yang-Mills theory on a cylinder,
in which the theory already reduces to a quantum mechanics problem.
By construction this prescription works only in two dimensions.
The quenched super gauge theory, being equivalent to the original
theory in the large $N$ limit, is an one dimensional $N=1$ supersymmetric
multi-matrix model. There are two Hermitian bosonic matrices, two Hermitian
fermionic matrices, or equivalently one complex bosonic matrix and one
complex fermionic matrix. This matrix model is interesting in its own right.
It is interesting to study its $1/N$ corrections, even these may have nothing
to do with $1/N$ corrections in the original super gauge theory.
In the similar spirit, one may try to generalize this supersymmetric
matrix model in various ways, and to ask the question whether a doubling
scaling limit exists, and if it does, what kind of string theory
it describes.

We shall not try to solve the quenched model directly in this
paper. The main result of this paper perhaps is the construction of
a solvable model by extending the supersymmetric quenched model to
include one more complex fermionic matrix.  There is still one complex
bosonic matrix. The bosonic part of the action
differs from the $N=1$ model by a potential term. The $N=1$ supersymmetry
in the extended model
can be explained as Parisi-Sourlas supersymmetry \ps.
Since such a model can be reduced to a Gaussian model together with
a pair of stochastic equations, Green's functions of bosonic
matrices are calculable in principle, once the first order stochastic equations
are solved. It is conceivable that further understanding of the extended
model will shed light on understanding of the quenched model.

Since the quenching prescription was proposed more than ten years ago,
it is appropriate to review the essential ingredients first. We shall
do this in the next section. We then proceed in sect.3 to discuss the
prescription for two dimensional gauge theories by starting with the
pure Yang-Mills theory defined on a cylinder. We shall argue that by
promoting the combination of momentum matrix and the spatial component
of the gauge field to a dynamic matrix, the average over momenta is
automatically done. Also, it is necessary to rescale the coupling constant.
The consistency of this prescription is checked in the case of an adjoint
matter coupled to the gauge field. The supersymmetric quenched model
is introduced in sect.4. We do not know how to solve this model yet.
Then in sect.5 we extend this model to obtain an exactly solvable model.
This model exhibits Parisi-Sourlas supersymmetry, and associated
first order stochastic equations can be integrated. The last section
is devoted to a discussion.

\newsec{A Brief Review of the Quenching Prescription}

To illustrate the idea of quenching, consider a two dimensional Hermitian
matrix $M(x,t)$ with the following action
\eqn\hermi{S(M)=\int d^2x\tr\left({1\over 2}(\p_t M)^2-{1\over 2}(\p_xM)^2
+{g\over\sqrt{N}}M^3\right),}
where the coupling constant $g$ is held fixed in the limit $N\rightarrow
\infty$.

Instead of quenching all spacetime momenta as in \refs{\bhn, \parisi,\gk},
only the spatial momentum is to be quenched in this paper \nkw. The
prescription is to replace the spatial dependent matrix field $M(x,t)$
by $\exp(iPx)M(t)\exp(-iPx)$, here $P$ is a diagonal matrix with eigenvalues
$p_i$. The derivative $\p_xM$ is replaced by $i[P,M]$ and the quenched
action reads
\eqn\quench{S=a\int dt\tr\left({1\over 2}(\p_tM)^2+{1\over 2}[P,M]^2
+{g\over\sqrt{N}}M^3\right),}
where $a=2\pi/\Lambda$ is the ultraviolet cut-off whose utility we will
see shortly, $M$ in the above action depends only on $t$. The propagator
derived from \quench\ is
\eqn\prop{\langle M_{ij}(t)M_{lk}(t')\rangle =-{i\over a}\delta_{ik}\delta_{jl}
(\p_t^2+p_{ij}^2)^{-1}\delta(t-t'),}
where $p_{ij}=p_i-p_j$, the momentum carried by $M_{ij}$. We thus see
that this propagator is almost the same as $\langle M_{ij}(p,t)M_{lk}(q,t')
\rangle$ in the original model if one lets $p=p_{ij}$, except for lacking
of the delta function factor $\delta(p+q)$.

We now show that any planar vacuum diagram in the original model is recovered
by averaging the corresponding diagram in the quenched model with an integral
\eqn\intfa{a^{-1}\prod_{i=1}^N\int^{\Lambda/2}_{-\Lambda/2}{adp_i\over 2\pi},}
each integral in the above product is normalized to 1. To see this, consider
a planar diagram with $l$ loops, $n$ vertices and $p$ propagators.
With the standard double-line
representation, on assigns momentum $p_i$ the line with index $i$, then
momentum conservation is automatic in the quenched model. An example
of vacuum diagrams is drawn in the following figure.

\vskip1cm
\epsfxsize=130pt \hskip1.5cm \epsfbox{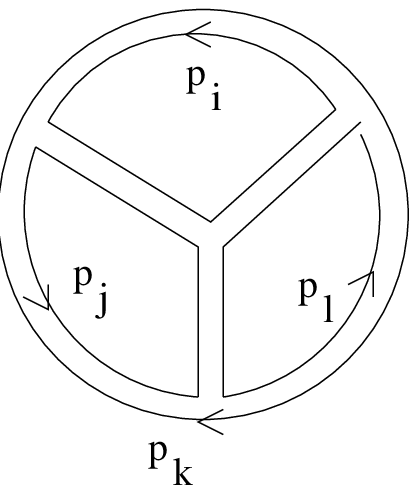}

The propagator is essentially the same as in the original model,
with an extra factor $a^{-1}$, so there is an additional factor $a^{-p}$
from all propagators. The $n$ vertices contribute a factor $(ag)^nN^{-n/2}$
and the contraction of matrix indices gives rise to a factor $N^{l+1}$.
So altogether there is a factor $a^{n-p}g^nN^{l+1-n/2}=
a^{n-p}g^nN^2$, where relations $3n=2p$ and $l-p+n=1$ are used. The factor
$N^2$ is the right one for a planar diagram. Now multiplying the result
by the integral factor \intfa, $l$ integrals in this factor are identified
with loop integrals in the original diagram, leaving a factor $a^l$
together $a^{-1}$ in \intfa. The rest integral factors normalize to 1. Thus,
the $a$ dependent factor is $a^{l-p+n-1}=1$, with the planar relation
$l-p+n=1$. This explains why a factor $a$ in the quenched action
is introduced. In the limit $N\rightarrow \infty$ all planar diagrams
are included, thus the free energy calculated with the quenching
prescription is the same as in the original model.

Calculating Green's functions requires a little modification. First,
according to the quenching prescription, $\tr M^n(x,t)=\tr M^n(t)$,
so Green's functions of these quantities will be independent of positions.
This is true in the large $N$ limit according to the factorization theorem.
Next, one would like to calculate the expectation value of quantities
such as $\tr [M(x,t)M(y,t')]$ or $\tr [M(p,t)M(q,t')]$. By the quenching
prescription
$$\langle \tr [M(p,t)M(q,t')]\rangle=(2\pi)^2\delta(p-p_{ij})\delta(q+p_{ij})
\langle M(t)_{ij}M(t')_{ji}\rangle,$$
where sum over indices is assumed. In the leading order, applying \prop\
and \intfa, one finds that the result is the same as in the original
model except for an additional factor $a^{-1}$. It is easy to see that
this factor exists for all connected planar diagrams contributing to this
Green's function. Similarly, when one considers a connected Green's function
of this type involving $n$ matrices, an additional factor $a^{-1}$ is to be
compensated.

\newsec{Quenched Yang-Mills Theory on a Cylinder}

\subsec{Pure Yang-Mills Theory on a Cylinder}

The pure Yang-Mills theory on a cylinder reduces to a quantum mechanics
problem \refs{\js, \dmp}, since the only dynamical degrees of freedom
are winding modes of the gauge field. Because the system itself already
effectively collapses to a ``point'' before quenching, it is desirable to
compare what obtained in the quenched model to the un-quenched model.

We start by showing how the pure Yang-Mills theory effectively reduces
to a quantum mechanical system. The standard action
\eqn\stand{S={1\over 2g^2}\int d^2x\tr(F_{01})^2}
can be cast into a first order form, upon introducing the canonical
momentum $\Pi^D$
\eqn\istord{\eqalign{S&=\int d^2x\tr\left(-{g^2\over 2}(\Pi^D)^2+
\Pi^D F_{01}\right)\cr
&=\int d^2x\tr\left(-{g^2\over 2}(\Pi^D)^2+\Pi^D\p_t A+A_0D_x\Pi^D\right),}}
where $A=A_1$. Integrating out $A_0$ in the path integral $\int [dA_0dA
d\Pi^D]\exp(iS)$ results in a delta function $\delta(D_x\Pi^D)$. This
forces $\Pi^D$ to satisfy $D_x\Pi^D=0$. The solution is given by
$$\Pi^D(x)=U^{-1}(x)\Pi^DU(x),\quad U(x)=P\exp(i\int^x_0 Adx),$$
where $\Pi^D$ is $\Pi^D(0)$, a matrix independent of $x$. Let the circumference
of the circle be $L$. By the periodic condition, $\Pi^D(L)=\Pi^D(0)=\Pi^D$,
one finds $[U,\Pi^D]=0$. Here $U$ is the holonomy around the circle
$U=U(L)$. Thus, upon substituting this solution into the path integral,
the delta function reduces to $\delta([U,\Pi^D])$ and the path
integral itself becomes
$$\int [dAd\Pi^D]\delta([U,\Pi^D])\exp\left(i\int dt\tr(-{g^2L\over 2}
(\Pi^D)^2+\Pi^D\int dx U(x)\p_tA(x)U^{-1}(x))\right).$$
It is easy to see that $\int dx U(x)\p_tA(x)U^{-1}(x))=-i\p_tUU^{-1}$,
from the definition of $U$. Thus, all degrees of freedom in $A(x)$ collapse
to that of $U$, and the above path integral finally reduces to a path
integral of a quantum mechanic system
\eqn\redu{\int [dUd\Pi^D]\delta([U,\Pi^D])\exp\left(i\int dt\tr(-{g^2L\over 2}
(\Pi^D)^2-i\Pi^D\p_tUU^{-1})\right),}
where the appropriate measure $[dU]$ is the Haar measure. Instead of
working with the holonomy, one can work with $D$ with the definition
$U=\exp(iDL)$. Using the following formula
$$-i\p_tUU^{-1}=L\int^1_0d\tau e^{iDL\tau}\p_tD e^{-iDL\tau}$$
in \redu\ and noting that $\exp(iDL\tau)$ effectively commutes with
$\Pi^D$, thanks to the delta function in the path integral, one finds
that these exponentials cancel upon taking trace. The delta function
$\delta([U,\Pi^D])$ can be replaced by another one $\delta([D,\Pi^D])$
and finally
\eqn\comp{
\int [dDd\Pi^D]\delta([D,\Pi^D])\exp\left(i\int dtL\tr(-{g^2\over 2}
(\Pi^D)^2+\Pi^D\p_tD)\right),}
where extra care need be exercised in definition of the measure $[dD]$.
One can define the measure as the usual flat one, and the nontrivial
factor coming from the Haar measure $[dU]$ can be absorbed into the
definition of the delta function. Eq.\comp\ is our final formula for
comparison with the quenched model.

Following the prescription given in the previous section, we use the
substitution
\eqn\subst{\Pi^D(x)=e^{iPx}\Pi^De^{-iPx},\quad
A_0(x)= e^{iPx}A_0e^{-iPx},\quad A(x)=e^{iPx}Ae^{-iPx}}
with a diagonal matrix $P$ whose entries take the form $p_i
=2\pi n_i/L$, $n_i$ is an integer. Furthermore, $D_x\Pi^D=i[D,\Pi^D]$,
where $D=P+A$. We shall see presently that $D$ is to be identified with
the previously introduced $D$ in \comp. The action \istord\ is replaced by
\eqn\repac{S=a\int dt\tr\left(-{g^2\over 2}(\Pi^D)^2+\Pi^D\p_t A+iA_0[D,\Pi^D]
\right).}
Again integration of $A_0$ results in constraint $[D,\Pi^D]=0$, and
the path integral becomes
\eqn\npath{\int[dDd\Pi^D]\delta([D,\Pi^D]\exp\left(i\int dt a
\tr(-{g^2\over 2}(\Pi^D)^2+\Pi^D\p_tA)\right).}
To see what quantity corresponds to $D$, consider the Wilson line
$U_{xy}=P\exp\int^y_xA(x)dx$. With the quenching substitution \subst,
$U_{xy}=\exp(iPx)\exp(iD(y-x))\exp(iPy)$, and in particular for the
Wilson line which wraps the circle once $U=\exp(iDL)$. We thus see
that $D$ is the same quantity as introduced in \comp. With this in mind,
the path integral \npath\ is seen to be almost the same as the un-quenched
version \comp. The differences between the two are that it is $A$ appears
in the action of the quenched version, and that the ultraviolet cut-off $a$
appears in the quenched action while it is the infrared cut-off $L$ appears
in the un-quenched action. The first difference is easily resolved by
replacing $\p_t A$ by $\p_tD$, since $P$ is supposed to be held fixed with
discrete eigenvalues.

Before resolving the second difference, we remind
ourselves that in the original proposal of \gk, the constraint that the
eigenvalues of $D$ coincide with those of $P$ is imposed. This amounts to
inserting the following factor
\eqn\const{\int [dU]\delta(D-UPU^{-1})\Delta(p_i),\quad
\Delta(p_i)=\prod_{i<j}(p_i-p_j)^2}
into the path integral.
We now argue that such constraint should not be imposed in our case of
a compact circle. As we have seen, an eigenvalue of $DL$ is the phase
of an eigenvalue of the holonomy $U$. With the above constraint, this
eigenvalue would be $2\pi n_i$ and the corresponding phase is trivial
in the holonomy. So the constraint \const\ would lead to a trivial
theory.

Now the second difference mentioned before between the two actions
is formally resolved by rescaling $\Pi^D\rightarrow {L\over a}\Pi^D$
and $g^2\rightarrow {a\over L}g^2$. The rescaling of $\Pi^D$ does not change
the quenched model, since it is just a matter of convention. The rescaling
of $g^2$ is a little disturbing. By viewing $D$, instead of $A$, as
the dynamic degrees of freedom, this rescaling appears necessary in
order to recover the original theory at the large $N$ limit. Such
modification is not absurd as it might appear. Note that except for the
Gauss law constraint $[D,\Pi^D]=0$, the action in \npath\ is quadratic,
so a straightforward perturbative argument as presented in the previous section
is lacking. In particular, averaging over $p_i$ with \intfa\ can not be
introduced directly. Finally, we would like to point out that
the ratio $L/a$ can be taken equal to $N$. The reason is the following.
The number of possible values of $p_i$ with both a ultraviolet cut-off and
an infrared cut-off is just $L/a$. When $P$ is absorbed into $D$, the
rank of $D$, $N$, counts effectively the number of possible values of $p_i$.
This is pointed out in \antal\ as well as in \sumit.

To check the consistency of our special prescription,
we shall consider the system of Yang-Mills coupled to an adjoint scalar
field in the next subsection.

\subsec{Yang-Mills Coupled to an Adjoint Scalar}

Let $\phi$ be an adjoint scalar, therefore a Hermitian matrix field for
a gauge group $U(N)$. In this subsection we are interested in studying the
coupled system with an action
\eqn\cpac{S={1\over 2g^2}\int d^2x\tr\left((F_{01})^2-(D_\mu\phi)^2\right),}
with definition of covariant derivatives $D_\mu\phi=\p_\mu\phi+i[A_\mu,\phi]$.
In the above we appropriately rescaled $\phi$ so that the action is
weighted by a factor $1/g^2$. This action can also be written in a first
order form
\eqn\firor{\eqalign{S&=\int d^2x\tr\left(-{g^2\over 2}[(\Pi^D)^2+(\Pi^\phi)^2]
+\Pi^DF_{01}+\Pi^\phi D_t\phi-{1\over 2g^2}(D_x\phi)^2\right)\cr
&=\int d^2x\tr\left(-{g^2\over 2}[(\Pi^D)^2+(\Pi^\phi)^2]+\Pi^D\p_tA
+\Pi^\phi \p_t\phi-{1\over 2g^2}(D_x\phi)^2\right)\cr
&+\int d^2x\tr A_0\left(D_x\Pi^D+i[\phi,\Pi^\phi]\right).}}
Again integrating out $A_0$ in the path integral imposes the Gauss law
$D_x\Pi^D+i[\phi,\Pi^\phi]=0$, which is solved by the following
expression
\eqn\slfm{\Pi^D(x)=U^{-1}(x)\Pi^DU(x)-i\int^x_0 dyU(x,y)[\phi(y),\Pi^\phi
(y)]U(y,x),}
where $\Pi^D=\Pi^D(0)$, and $U(x)$ is defined as in the last subsection, and
$U(x,y)=P\exp(i\int^y_xA(x)dx)$. Now if one puts the system on a circle
of circumference $L$, the periodic boundary condition for $\Pi^D(x)$
no longer results in $[U,\Pi^D]=0$, because of the second term in \slfm.
Also, after substituting \slfm\ into the action \firor, the action
not only depends on $U=U(0,L)$, but also on $U(x,y)$. Therefore,
in the quenched
model, the eigenvalues of $D$, as defined in the previous subsection is no
longer restricted to a circle, since $U(x,y)$ is not well defined on this
circle. Thus, unlike in the pure Yang-Mills theory, the momentum as part
of an eigenvalue of $D$ will enter in the story in an essential way.

The solution \slfm\ of the constraint can be simplified by using
gauge transformation
$$\eqalign{&\phi(x)\rightarrow U^{-1}(x)\phi(x)U(x),\quad
\Pi^\phi(x))\rightarrow U^{-1}(x)\Pi^\phi(x))U(x),\cr
&\Pi^D(x)\rightarrow U^{-1}(x)\Pi^D(x)U(x), \quad A(x)\rightarrow
U^{-1}(x)A(x)U(x)-iU^{-1}(x)\p_xU(x).}$$
The action \firor\ is invariant under this transformation, since the additional
term resulting from the transformation is proportional to the constraint,
therefore vanishes. The solution \slfm\ after subject to this transformation
reads
\eqn\asl{\Pi^D(x)=\Pi^D-i\int^x_0 dy[\phi(y),\Pi^\phi(y)].}
Since $U(x)$ is not necessarily periodic, the periodic boundary conditions
for new fields are twisted by $U$, for example $\Pi^D(L)=U\Pi^D(0)U^{-1}$.
With the help of the above solution, this condition translates into
\eqn\resi{[U,\Pi^D]=-i\int^L_0 dy[\phi(y),\Pi^\phi(y)]].}
One can substitute \asl\ into action \firor. Once again, one would obtain
a term
$$-{g^2L\over 2}\int dt\tr(\Pi^D)^2$$
in which the infrared cut-off $L$ appears explicitly. This suggests that the
quenching prescription introduced in the previous subsection also works in
this model. To further confirm this, we suggest an unconventional perturbation
scheme in which the whole term
$$-{1\over 2g^2}\int d^2x\tr\left(D_x\phi\right)^2$$
is treated as a perturbation term. Certainly there are other terms resulting
from solving the constraint. Let us focus on the above term. To normalize
the kinetic terms in \firor, one has to rescale $\phi\rightarrow g\phi$,
$\Pi^\phi\rightarrow \Pi^\phi/g$, $D_x\rightarrow gD_x$. Thus, the above
perturbation becomes
\eqn\pert{-{g^2\over 2}\int dt{1\over L}\sum_n \tr(D_x\phi)^2,}
where we replaced the integral over the spatial dimension by a sum over
discrete momentum modes. Accordingly, the quantity $\tr(D_x\phi)^2$
should be viewed as the Fourier transform of the original term. Essentially,
there is no ultraviolet cut-off involved in this perturbation theory.
$D_x\phi$ is effectively replaced by $ip\phi+i[a,\phi]$ where $a$ is the
constant mode of the gauge field. For a detailed discussion we refer to \shif\
where an adjoint fermion coupled to the gauge field is discussed.
Now one can replace the sum over momentum modes in \pert\ by a trace,
thus the perturbation \pert\ becomes a four point vertex and the combination
of $p+A$ becomes an independent field.

Finally, we are in a position to propose our quenching prescription. First,
as usual,
\eqn\asua{\eqalign{A(x)&=e^{iPx}Ae^{-iPx},\quad \Pi^D(x)=e^{iPx}\Pi^De^{-iPx},
\cr
\phi(x)&=e^{iPx}\phi e^{-iPx},\quad \Pi^\phi(x)=e^{iPx} \Pi^\phi(x)e^{-iPx}.}}
With these substitutes, the action in \firor\ becomes
\eqn\npres{S=L\int dt\tr\left(-{g^2\over 2}[(\Pi^D)^2+(\Pi^\phi)^2]
+\Pi^D\p_tD+\Pi^\phi\p_t\phi+{1\over 2g^2}[D,\phi]^2\right),}
with constraint
\eqn\ncons{[D,\Pi^D]+[\phi,\Pi^\phi]=0.}
It is easy to see that the above constraint is a consequence of the old
one with prescription \asua. What is new here is that the usual ultraviolet
cut-off $a$ is replaced by the infrared cut-off $L$, and $D$ is promoted
to a dynamic variable so the average over momentum is not necessary. This
is also why a ultraviolet cut-off is not necessary, since the propagator of
$\phi$ does not involve $P$. The necessity of an infrared cut-off is best
shown in study of the pure Yang-Mills theory. To see that this prescription
is consistent, we note that any gauge invariant quantity, such as
$$\tr[\phi(x)U(x,y)\phi(y)U(y,x)]=\tr \phi\exp(iD(y-x))\phi\exp(iD(x-y)),$$
contains $P$ through $D$. The prescription works
only for gauge theories, since in a matrix model with only global symmetry,
$P$ appears explicit in all correlation functions.

Before closing this section, we mention once again that the Gross-Kitazawa
constraint \const\ should not be imposed in this model on a compact circle
too. And the eigenvalues of $D$, unlike in the pure gauge theory, should
not be restricted to live on a circle.

\newsec{The Quenched Supersymmetric Yang-Mills Theory}

\subsec{The Quenched Model}

In this section, we shall introduce the quenched supersymmetric gauge
theory. First, let us introduce the 2D super-gauge theory. The vector
super-multiplet
in two dimensions consists of a vector field, a scalar field and a Majorana
fermion. The scalar field is needed, since the vector field in two dimensions
does not have dynamical degrees of freedom. The  action
\eqn\action{S={1\over g^2}\int d^2x\tr\left(-{1\over 4}F_{\mu\nu}^2
-{1\over 2}(D_\mu\phi)^2-i\lambda\bar{\sigma}^\mu D_\mu\lambda
-\lambda\sigma^3[\phi, \lambda]\right),}
is invariant under the SUSY transformation:
\eqn\supertr{\eqalign{\delta A_\mu&=-2i\lambda\bar{\sigma}_\mu\epsilon,\cr
\delta\phi&=2i\lambda\sigma^3\epsilon,\cr
\delta\lambda&=\sigma^1F_{01}\epsilon -\sigma^\mu\sigma^3D_\mu\phi \epsilon.}}
We use notations of \wb, $\bar{\sigma}_\mu=(1,-\sigma^1)$. The constant
$g^2N$ is held fixed in the limit
$N\rightarrow \infty$. Here for simplicity we consider the gauge group $U(N)$.

It is necessary to fix a gauge in the Hamiltonian
formalism. An arbitrary gauge will spoil supersymmetry, thus many advantages
brought about by SUSY may disappear. As was shown in \miao, there is
a gauge in which half of SUSY is broken, but half of SUSY survives.
In this gauge, $A_0+\phi=0$. Thus we are left with two bosonic fields
$A=A_1$ and $\phi$, and two fermionic fields $\lambda_\alpha$. The
conjugate momenta are
\eqn\momen{\eqalign{\Pi^A&={1\over g^2}F_{01}={1\over g^2}\left(\p_0A+D_x\phi
\right),\cr
\Pi^\phi&={1\over g^2}\p_0\phi,\quad \Pi_\lambda=i{1\over g^2} \lambda.}}
with the standard commutation relations
\eqn\commu{\eqalign{[\Pi^A_{ij}(x),A_{lk}(y)]&=-i\delta_{ik}\delta_{jl}
\delta(x-y),\quad
[\Pi^\phi_{ij}(x),\phi_{lk}(y)]=-i\delta_{ik}\delta_{jl}\delta(x-y),\cr
\{\lambda^{ij}_\alpha(x),\lambda^{lk}_\beta(y)\}&={g^2\over 2}\delta_{ik}
\delta_{jl}\delta_{\alpha\beta}\delta(x-y).}}

Now the Hamiltonian and the unbroken super-charge $Q$ are, respectively
\eqn\hamil{\eqalign{H&=g^2\int dx\tr\left({1\over 2}(\Pi^A)^2+
{1\over 2}(\Pi^\phi)^2\right)-\int dx\tr\Pi^AD_x\phi\cr
&+{1\over g^2}\int dx\tr\left({1\over 2}(D_x\phi)^2-i\lambda \sigma^1
D_x\lambda +\lambda(\sigma^3-1)[\phi,\lambda]\right),}}
and
\eqn\scharge{Q=\int dx\tr\left(\lambda_1\Pi^\phi+\lambda_2(\Pi^A
-{1\over g^2}D_x\phi)\right).}
The relation between these two quantities is $H=\{Q,Q\}$.

Now we are in a position to introduce the quenched model. The model is
obtained by replacing every matrix field $F(x)$ with $\exp(iPx)F
\exp(-iPx)$, where $F$ is independent of $x$ and $P$ is a real diagonal
matrix, as in the previous section. The covariant derivative $D_xF(x)$ is
then replaced by
$\exp(iPx)i[D,F]\exp(-iPx)$, where $D=P+A$. On a circle, the
eigen-values of $P$ take discrete values $2\pi n/L$, $n$ is an integer.
With these substitutions, the Hamiltonian and the super-charge are
truncated to a point
\eqn\qhamil{\eqalign{H&={g^2L\over 2} \tr\left((\Pi^D)^2+(\Pi^\phi)^2
\right)-iL\tr\Pi^D[D,\phi]\cr
&+{L\over g^2}\tr\left(-{1\over 2}[D,\phi]^2+\lambda \sigma^1 [D,\lambda]
+\lambda(\sigma^3-1)[\phi,\lambda]\right).}}
and
\eqn\qschar{Q=L\tr\left(\lambda_1\Pi^\phi+\lambda_2(\Pi^D
-{1\over g^2}i[D,\phi])\right),}
where we replaced $\Pi^A$ by $\Pi^D$.
Still these quantities are not convenient to work with, since the commutation
relations are given by \commu\ with $\delta(x-y)$ replaced by $1/L$.
Note that here our prescription is different from the usual one, in that
the delta function is not regularized by $1/a$ with a short-distance cut-off
$a$. The argument is the same as we presented in the last section.
The commutators are simplified by substitutions
$$\eqalign{&D\rightarrow {g\over \sqrt{L}}D,\quad \phi\rightarrow {g\over
\sqrt{L}}\phi,\quad \lambda\rightarrow {g\over\sqrt{L}}\lambda ,\cr
&\Pi^D\rightarrow {1\over g\sqrt{L}}\Pi^D,\quad \Pi^\phi\rightarrow
{1\over g\sqrt{L}}\Pi^\phi,}$$
Explicitly
\eqn\qcommu{\eqalign{[\Pi^D_{ij},D_{lk}]&=-i\delta_{ik}\delta_{jl},\quad
[\Pi^\phi_{ij},\phi_{lk}]=-i\delta_{ik}\delta_{jl},\cr
\{\lambda^{ij}_\alpha,\lambda^{lk}_\beta\}&={1\over 2}\delta_{ik}
\delta_{jl}\delta_{\alpha\beta}.}}
The super-charge reads
\eqn\qscharge{Q=\tr\left(\lambda_1\Pi^\phi+\lambda_2(\Pi^D
-il[D,\phi])\right),}
where the parameter $l=g/\sqrt{L}$. The final ingredient is
the gauge transformation
\eqn\gtrans{\eqalign{&D\rightarrow UDU^{-1},\quad \phi\rightarrow U\phi U^{-1},
\quad \lambda\rightarrow U\lambda U^{-1},\cr
&\Pi^D\rightarrow U\Pi^DU^{-1}, \quad \Pi^\phi\rightarrow U\Pi^\phi U^{-1},}}
where the original gauge transformation parameter $U(x)$ is also
replaced by
$$\exp(iPx)U\exp(-iPx).$$
The above gauge transformation is
generated by the generator
\eqn\genera{G=i[D,\Pi^D]+i[\phi,\Pi^\phi]+2\lambda^2.}
A physical state is annihilated by $G$.

\subsec{The Superspace Formulation}
The super-charge \qscharge\ can be written in a more symmetric fashion,
if one introduces
\eqn\newn{\eqalign{\lambda&=\lambda_1-i\lambda_2,\quad \bar{\lambda}=
\lambda_1+i\lambda_2,\cr
M&=\phi -iD,\quad \overline{M}=\phi +iD,\cr
A&=\Pi^M-{il\over 4}[\overline{M},M],\quad
\bar{A}=\Pi^{\overline{M}}+{il\over 4}[\overline{M},M],}}
then
\eqn\simp{Q=\tr(\lambda A+\bar{\lambda}\bar{A}),}
with corresponding action
\eqn\redac{\eqalign{S&=\int dt\tr[{1\over 2}\p_tM\p_t\overline{M}
-{il\over 4}[\overline{M},M] \p_t(\overline{M}-M)+i\bar{\lambda}\p_t\lambda\cr
&- {l\over 2}\lambda [\overline{M},\lambda]-  {l\over 2}\bar{\lambda}[M,
\bar{\lambda}]+{l\over 2}\bar{\lambda}[M+\overline{M},\lambda]].}}
This action is invariant under the following SUSY transformation
\eqn\susyt{\eqalign{\delta M&=-2i\epsilon\lambda,\quad
 \delta \overline{M}=-2i\epsilon\bar{\lambda},\cr
\delta\lambda&=\epsilon\p_t M, \quad \delta\bar{\lambda}=\epsilon\p_t
\overline{M}.}}

Since the above transformation has a simple form, one would guess there
might be a simple superspace formulation. Indeed there is. Introduce a real
fermionic coordinate $\theta$ as the superpartner of $t$. Let $D=\p_\theta
-i\theta\p_t$ be the super-covariant derivative. The super-charge
commuting with $D$ is $Q=\p_\theta +i\theta\p_t$. There is $\{Q,Q\}
=2i\p_t=2H$. Introduce the following super-fields
\eqn\sufd{\Phi=M+i\theta\lambda, \quad \overline{\Phi}=\overline{M}
+i\theta\bar{\lambda}.}
It is easy to see that the free part of the action \redac\ is
$$-{1\over 2}\int dtd\theta \tr D\overline{\Phi}D^2\Phi=
\int dt\tr{1\over 2}\left(\p_tM\p_t \overline{M}+i\bar{\lambda}
\p_t\lambda\right),$$
where there is an extra factor $1/2$ for the fermionic part, which is
due to a rescaling $\lambda\rightarrow \lambda/\sqrt{2}$.
The cubic terms in \redac\ can also be sorted out easily. We simply
write down the whole action in terms of the super-fields
\eqn\whac{S=\int dtd\theta \tr\left(-{1\over 2}D\overline{\Phi}D^2\Phi
+{l\over 4}[\overline{\Phi},\Phi]D(\overline{\Phi}-\Phi)\right).}

\newsec{An Extended Model With Parisi-Sourlas Supersymmetry}

We have tried to solve the quenched model introduced in the last section,
for example, by constructing some conserved quantities. So far we have not
succeeded in tackling this model directly. The reason why any simple minded
method of constructing conserved quantities does not work is that all
$A$ and $\bar{A}$ introduced in \newn\ do not commute. For example, a
commutator of two matrix elements of $A$ depends on $\overline{M}$.

In this section, we suggest to study an exactly solvable model with
one more complex fermion. This model is very similar to the $N=1$ model
introduced in the previous section. We believe that better understanding
of this model shall shed light to the $N=1$ quenched model.

The supersymmetry to be introduced here is Parisi-Sourlas supersymmetry.
Therefore it is convenient to follow the line of the original papers
\ps\ to introduce this model. We start with an action obtained by adding
a potential term
$${l^2\over 8}\tr[\overline{M},M]^2,$$
to the bosonic part of \redac. This term is positive definite,
since $[\overline{M},M]$ is Hermitian. This
means that a negative definite term is added to the quenched Hamiltonian.
The so obtained bosonic action is
\eqn\defac{S(M,\overline{M})=\int dt\tr[{1\over 2}(\p_tM-{il\over
2}[\overline{M},
M])(\p_t\overline{M}+{il\over 2}[\overline{M},M])].}
Note that this action is a complete square.

Define the complex Gaussian fields
\eqn\langi{\eta=\p_tM-{il\over 2}[\overline{M},M], \quad
\bar{\eta}=\p_t\overline{M}+{il\over 2}[\overline{M},M],}
with the measure
\eqn\meas{\int [d\eta d\bar{\eta}]e^{{i\over 2}\int dt\tr\eta\bar{\eta}}.}
The Green's functions of $M$ and $\overline{M}$ are defined through
solution to \langi\ together with the above Gaussian path integral.
Denote the solution (with appropriate initial conditions) of \langi\
by $M=M(\eta,\bar{\eta})$, $\overline{M}=\overline{M}(\eta,\bar{\eta})$.
A Green's function is then
$$\eqalign{&\int [d\eta d\bar{\eta}]F[M(\eta,\bar{\eta}) ,\overline{M}
(\eta,\bar{\eta}]
e^{{i\over 2}\int dt\tr\eta\bar{\eta}}\cr
&=\int [dMd\overline{M}]det[{\p(\eta,\bar{\eta})\over\p (M,\overline{M})}]
F[M,\overline{M}]e^{iS},}$$
where in the second line we changed the integral variables from the $\eta$'s
to the $M$'s, and $S$ is the action given by \defac. The Jacobian, following
relations \langi, is given by
\eqn\jacob{{\p(\eta,\bar{\eta})\over\p (M,\overline{M})}
=\left( \matrix{\p_t-{il\over 2}[\overline{M},\cdot]&-{il\over 2}[\cdot, M]\cr
-{il\over 2}[\cdot, \overline{M}]&\p_t-{il\over 2}[M,\cdot]}\right) .}
Its determinant can be expressed as a fermionic path integral. Let
$\Psi=[\lambda,\psi]^t$, then
$$det[{\p(\eta,\bar{\eta})\over\p (M,\overline{M})}]=
\int [d\Psi d\overline{\Psi}]\exp\left(iS(\Psi,\overline{\Psi})\right)$$
with
\eqn\ferac{\eqalign{S(\Psi,\overline{\Psi})&=i\int dt \tr(\psi,\bar{\psi})
{\p(\eta,\bar{\eta})\over\p (M,\overline{M})}(\lambda,\bar{\lambda})^t\cr
&=
\int dt\tr\left(i\psi\p_t\lambda
+i\bar{\psi}\p_t\bar{\lambda}+{l\over 2}(\psi[\overline{M},\lambda]+\bar{\psi}
[M,\bar{\lambda}]-\psi[\overline{M},\bar{\lambda}]-\bar{\psi}[M,\lambda])
\right).}}
Note that the structure of this action is almost the same as the fermionic
part of action \redac, except here there are two complex fermionic matrices.
(Formally if $\psi\rightarrow \bar{\lambda}$ and $\bar{\psi}\rightarrow
\lambda$, then the above action is the same as in \redac.)
Now a Green's function of $M$ and $\overline{M}$ can be written as
\eqn\green{\langle F[M, \overline{M}]\rangle=   \int[dM\overline{M}d\Psi
d\overline{\Psi}]F[M, \overline{M}]
e^{i[S(M, \overline{M})+S(\Psi, \overline{\Psi})]}.}

As always, the combined action $ S(M, \overline{M})+S(\Psi, \overline{\Psi})$
possesses supersymmetry, Parisi-Sourlas supersymmetry. Unlike in the simplest
cases, the supersymmetry parameter is a real Grassmanian:
\eqn\pssup{\eqalign{\delta M&=-2i\epsilon\lambda,\quad \delta\overline{M}
=-2i\epsilon\bar{\lambda},\cr
\delta\psi&=\epsilon\left(\p_t\overline{M}+{il\over 2}[\overline{M},M]
\right),\cr
\delta\bar{\psi}&=\epsilon\left(\p_t M-{il\over 2}[\overline{M},M]
\right),}}
$\lambda$ and $\bar{\lambda}$ are invariant under this supersymmetry. We
suspect that there is one more supersymmetry, perhaps a nonlinear one,
but have not been able to sort it out. The supersymmetry \pssup\
itself does not generate the Hamiltonian, precisely because the $\lambda$'s
are invariant under it. To be able to generate the Hamiltonian, one perhaps
need another supersymmetry.

The first order nonlinear differential equations in \langi\ are integrable.
Observe that the ``friction'' term $il/2[\overline{M},M]$ is anti-Hermitian,
so the Hermitian part of $M$ satisfies a linear equation. Use the same
notation as in \newn, denote this Hermitian part by $\phi$ and the
anti-Hermitian part by $D$. A general solution for $\phi$ is
\eqn\lins{\phi(t)=\int_{-\infty}^tdt'\hbox{Re}\eta(t')+\phi(-\infty),}
where $\hbox{Re}\eta$ denotes the Hermitian part of $\eta$, $\phi(-\infty)$
is the value of $\phi$ at the infinite past. Now $D$ satisfies the following
equation
\eqn\nlang{\p_tD-il[\phi, D]=-\hbox{Im}\eta,}
with $\hbox{Im}\eta$ being the anti-Hermitian part of $\eta$. Since $\phi(t)$
is known as in \lins, the solution to the above equation can be expressed
in terms of $\phi$,
\eqn\linss{D(t)=-\int_{-\infty}^tdt'U^{-1}(t',t)\hbox{Im}\eta(t')U(t',t)
+D(-\infty),}
where
$$U(t',t)=P\exp(-il\int_{t'}^t\phi(\tau)d\tau).$$
That $\phi$ appears in the above path ordered integral is not surprising.
Recall that in the quenched model, $\phi$ is just $-A_0$ in the gauge
in which one supersymmetry is preserved. Thus, $U(t',t)$ is just
the Wilson line along the time direction.

Now any Green's function involving only matrices $\phi$ and $D$, or
equivalently $M$ and $\overline{M}$, can be calculated using \lins\ and
\linss\ together with the Gaussian path integral \meas. We shall not
try to develop a systematic technique to do this in the present paper,
but shall leave it for future work. Since the $\eta$'s are Gaussian fields,
it is possible to treat the large $N$ problem with the help of the
master field, for these are the so-called free variables. Some Green's
functions involving fermionic matrices can also be calculated, using
the Ward identities associated with SUSY in \pssup. For instance, starting
with
$$\langle\delta_\epsilon\tr[\psi M]\rangle=0$$
one derives
\eqn\fcorr{\langle\tr[\psi\lambda]\rangle= {i\over 2}\langle\tr[M
(\p_t\overline{M}+{il\over 2}[\overline{M},M])]\rangle.}
This Ward identity holds for finite $N$. There is no anomalous term,
since SUSY is not spontaneously broken. This is guaranteed by the
fact that the solution to the Langevin equations \langi\ is unique for
given initial data.

Finally, a technical issue to be addressed is the projection to the
singlet sector. The solution is not difficult and again we leave it to
future work.

\newsec{Discussion}

We have shown that the two dimensional supersymmetric Yang-Mills theory
truncates to an one dimensional supersymmetric matrix model in the
large $N$ limit. Although the quenched model may have nothing to do with
the original SYM theory beyond the large $N$ limit, it is tempting
to treat itself as an interesting matrix model, and to ask various
questions as often arise in dealing with a matrix model. Before the
large $N$ problem becomes tractable, asking these questions remain
un-practical.

The extended matrix model studied in sect.5, exhibiting
Parisi-Sourlas supersymmetry, can be solved exactly. This model is very
similar to the quenched model, so further study will undoubtedly shed light on
understanding of the quenched model. Some mechanism is needed to
truncate the number of fermionic matrices, in order to come down to the
quenched model. We have studied the quenched $N=2$ two dimensional
SYM theory, and the resulting theory contains the same number of fermions
as in the extended model. There is $N=2$ supersymmetry and an additional
bosonic Hermitian matrix. We intend to publish some results concerning
this model elsewhere. For now, we just mention that the $N=2$ 2D
SYM theory is the dimensional reduction of the $N=1$ 4D SYM theory,
and the latter is shown to possess the Nicolai mapping \nicolai.
In a sense, the Nicolai mapping is nothing but a generalized Parisi-Sourlas
mapping. Therefore, we suspect that this model might be interconnected
with the two models studied in this paper.

It is our hope that some higher dimensional string theory will eventually
emerge from these supersymmetric matrix models. With multi-matrices at
hand, it is possible to generate more than one dimensions. Also, so far
only supersymmetry can help in overcoming the so-called $c=1$ barrier,
and in eliminating problems such as tachyon. We hope to return to models
introduced in this paper in near future.

\noindent {\bf Acknowledgments}

We would like to thank M. Douglas for initial encouragement, and S. Das
and A. Jevicki for very useful discussions. This work was supported
by DOE grant DE-FG02-91ER40688-Task A.

\listrefs

\end